
\magnification=1200

\nopagenumbers
\headline={\ifnum\pageno>1\hss -- \tenrm\folio -- \hss
\else\hfil\fi}
\baselineskip 14pt
\centerline{\bf Domain Walls in Non-Equilibrium Systems and the Emergence}
\centerline{\bf of Persistent Patterns}
\vskip .4truein
\baselineskip 16pt
\centerline{\sl Aric Hagberg and Ehud Meron}
\bigskip
\centerline{Arizona Center for Mathematical Sciences}
\centerline{and Program in Applied Mathematics}
\centerline{University of Arizona}
\centerline{Tucson, AZ 85721}
\vskip 1.truein
\baselineskip 20pt
\noindent
{\sl Abstract:}
\bigskip

Domain walls in equilibrium phase transitions propagate
in a preferred direction so as to minimize the free energy of the system.
As a result, initial spatio-temporal patterns ultimately decay toward
uniform states. The absence of a variational principle far from equilibrium
allows the coexistence of domain walls propagating in any direction.
As a consequence, {\it persistent} patterns may emerge.
We study this mechanism of pattern formation using a non-variational extension
of Landau's model for second order phase transitions.
\vskip 1.5truein
\noindent
PACS numbers: 05.70.Fh, 42.65.Pc, 47.20.Ky, 82.20Mj \hfil

\vfil
\eject
Second order phase transitions, such as in ferromagnets or
liquid-vapor systems, are manifestations of spontaneous symmetry
breaking occurring near thermal equilibrium.
The coexistence of broken symmetry states beyond the transition
point gives rise to spatial patterns consisting of domain walls or
fronts separating regions of different phase.
The dynamics of domain walls near thermal equilibrium,
are dictated by a variational principle, namely, the minimization of the
free energy. As a consequence, initial spatio-temporal patterns ultimately
decay toward the stationary homogeneous states of lowest free energy.
Front structures are commonly observed in
far-from-equilibrium systems as well. Walls separating conductive
and convective states in binary mixtures [1], excited and recovery regions
in autocatalytic chemical reactions [2], or different phase-locked states
in parametrically forced surface waves [3], are a few examples. Unlike
equilibrium systems, however,
no general variation principle, like the minimization of free energy,
applies for systems maintained far from equilibrium.
The possible outcome, as emphasized recently [4-7], is the appearance of
localized structures and, more generally, persistent spatio-temporal
patterns.

In this paper we further elaborate on pattern formation as a non-variational
effect. We consider spatio-temporal patterns involving domain walls and
show that multiplicity of stable front solutions may give rise to
persistent patterns. As a model system we choose to study a non-variational
extension of the Landau-Ginzburg model for a scalar
order parameter. The extended system takes the form of coupled reaction
diffusion equations that have extensively been studied in the context of
chemical and biological patterns [8-13]. Some of the results to be
described here have already been obtained in that context before,
particularly by Rinzel and Terman [12]. We rederive these results and
present them here in a way that best illustrates the
different point of view we put forward in this paper.

Consider a variational system whose free energy (Liapunov functional) is
given by
$${\cal F}=\int \big[{\cal U}(\phi,h)
+\phi_x^2/2\bigr]dx,\eqno(1)$$
where ${\cal U}(\phi,h)=-\phi^2/2+\phi^4/4+h\phi$,~
$\phi(x,t)$ is a scalar order parameter, $h$ is a constant bias field
and the subscript $x$ denotes the spatial partial derivative.
For $h$ values in the range $-2/(3\sqrt 3)<h<2/(3\sqrt 3)$ the free energy
density, ${\cal U}$,  has a double-well form. The two wells
correspond to stationary homogeneous
states, characterized by order-parameter values $\phi_-(h)$  and $\phi_+(h)$
that solve the cubic equation $\phi^3-\phi+h=0$.
The relaxation toward any of these stationary
states is governed by the equation $\phi_t=-\delta{\cal F}/\delta\phi$,
or
$$\phi_t=f(\phi,h)
+\phi_{xx}~~~~~~~~~~f(\phi,h)=\phi-\phi^3-h.\eqno(2)$$
Front solutions, $\phi=\phi(\chi)$ where $\chi=x-ct$, of (2) [14]
propagate in a preferred direction dictated by the minimization of ${\cal F}$.
The speed of a front connecting $\phi_+(h)$ at $\chi=-\infty$ to $\phi_-(h)$
at $\chi=\infty$ is given by
$$c=\mu(h)
=\alpha[{\cal U}(\phi_-)-{\cal U}(\phi_+)],\eqno(3)$$
where $\alpha(h)=1/\int_{-\infty}^\infty {\phi^\prime(h)}^2 d\chi$ is
positive [11]. For negative $h$ values, ${\cal U}(\phi_-)>{\cal U}(\phi_+)$
and the front moves in the positive $x$ direction ($c>0$) so as to increase
that part of the system having lower energy. When $h$ is positive
the front propagates toward negative $x$ values  ($c<0$). Notice that
(2) remains invariant under the transformation
$x\to -x$. Thus, in addition to a front
connecting $\phi_+(h)$ to $\phi_-(h)$ and propagating, say, at positive speed
$c$, there exists a symmetric front connecting $\phi_-(h)$ to $\phi_+(h)$
and propagating at a negative speed $-c$. In the following we refer to
these symmetric solutions as representing the same type of front solution.

Imagine now that $h$ is not constant but, instead, a second field
$h=h(x,t)$ coupled to $\phi=\phi(x,t)$.
A variety of physical, chemical and biological systems fall
in that category. Bistable optical
systems [15], crystal growth [16], autocatalytic chemical reactions [2,8],
and predator-prey systems [17],
are a few examples. For the present purposes it is sufficient
to consider the simplest case where $h(x,t)$ is a diffusive field that responds
linearly to changes in the order parameter $\phi(x,t)$. More specifically
we assume the form [8]
$$h_t=\epsilon g(\phi,h)
+\delta h_{xx}~~~~~~~~~~g(\phi,h)=\phi-a_0-a_1 h,
\eqno(4)$$
where $\epsilon>0$ is the ratio, $\tau_\phi/\tau_h$, between the time
scales associated with the two fields $\phi$ and $h$,
$\delta=D_h/D_\phi$ is the ratio of diffusion constants, and
the coefficient $a_1$ is positive.
The combined system (2) and (4) (denoted hereafter by (2+4)) is no longer
variational.
Yet, it resembles the original system (2) in having, for a proper choice of
$a_0$ and $a_1$, three stationary homogeneous solutions of which two are
stable. The stable solutions correspond to the intersection points of the
nullcline $g(\phi,h)=0$ with the
branches $\phi=\phi_\pm(h)$ and are denoted here by
$(\phi_\pm,h_\pm)$ (see Figs. 1). Unlike the variational
system, however, the two stable states can be connected
by more than one type of front solutions when $\epsilon$ is sufficiently
small [10-13].
Preparing the system at the lower state $(\phi_-,h_-)$ and perturbing it
locally so as to induce a transition to the upper branch $\phi=\phi_+(h)$,
yields a front propagating to the right: $(\phi(\chi),h(\chi))\to
(\phi_\pm,h_\pm)~~{\rm as}~~ \chi\to\mp\infty,~~c>0$.
If, on the other hand, the initial state is the upper one, $(\phi_+,h_+)$,
a perturbation that induces a transition to the lower branch
$\phi=\phi_-(h)$ yields a front connecting the same asymptotic states but
propagating to the left:
$(\phi(\chi),h(\chi))\to (\phi_\pm,h_\pm)~~{\rm as}~~ \chi\to\mp\infty,~~c<0$.
The two fronts are not
related by the symmetry $x\to -x$ and, therefore,
represent two different types of front solutions.

The multiplicity of front solutions and the symmetry
$x\to -x$ of (2) imply that
along with a front that transforms the lower state
$(\phi_-,h_-)$ to the upper state $(\phi_+,h_+$), there exists another
front (hereafter ``back'') propagating in
the {\it same direction} that transforms the upper state
back to the lower one. A combination of the two
may yield a persistent localized structure, provided there exists a
mechanism which binds the back to the front.

To study the emergence of such
a structure we consider the small $\epsilon$ regime, $\epsilon\ll 1$,
and assume a nondiffusive $h$ field, or
$\delta=0$ (allowing diffusion of $h$ will not affect the results
qualitatively as long as $\delta$ is not too large).
We then distinguish between front and back regions where $h$
barely changes, and outer regions where $\phi$ can be eliminated
adiabatically, $\phi=\phi_\pm(h)$ [8-10]. Imagine now a front transforming
the down state $(\phi_-,h_-)$ into the upper one $(\phi_+,h_+)$ and propagating
to the right. The front speed is determined by the local value of $h$:~
$c=\mu(h_-)>0$. A back that follows the front (so as to form a single
up-state domain) will be affected by the field
$h(\chi)$ that develops behind the front. This field, as we will show below,
provides the binding force the front exerts on the back.
In order to find that field we insert
$\phi=\phi_+(h)$ in (4) to obtain a closed equation for $h$, and use the
boundary conditions
$h(\chi_f^0)=h_-$ and $h(\chi)\to h_+$ as $\chi\to -\infty$, where
$\chi_f^0$ denotes the front position in a frame moving at speed $c$.
Solving for $h$ we find
$$h(\chi)=(h_--h_+)e^{\epsilon\kappa(\chi-\chi_f^0)}+h_+~~~~~~~~~~~~
\chi\le\chi_f^0,\eqno(5)$$
where $h_\pm=2(\pm 1-a_0)/(1+2a_1)$ and  $\kappa=(a_1+1/2)/c$.
In deriving (5) we used the linear
approximation $\phi_\pm(h)\approx\pm 1-h/2$ valid for small $\vert h\vert$.

For an up-state domain to become a localized structure of fixed size, the back
speed should be equal to the front speed $c$. We thus require
$\mu(h_b^0)=-c$, where $h_b^0=h(\chi_b^0)$ is the local value of $h$ at the
back. This relation determines $h_b^0$. Note that $h_b^0$ must be positive,
whereas $h_-$, the value of $h$ at the front, is negative.
Inserting $h=h_b^0$ and $\chi=\chi_b^0$ in (5) we find
for the size of the localized structure
$$\lambda=\chi_f^0-\chi_b^0={1\over \epsilon\kappa}
\ln\Bigl({h_+-h_-\over h_+-h_b^0}\Bigr).\eqno(6)$$
According to (6), a localized structure of fixed size exists for $h_+>h_b^0$
(for $h_+<h_b^0$ the front speed is larger than the back speed and the
up-state domain expands indefinitely). We will now show that this structure
is also stable to translational perturbations. To this end we represent the
back in the form
$$\phi_b(x,t)=\phi_b^0(\chi-\chi_b)+\epsilon\phi_b^1(\chi-\chi_b,\epsilon t),$$
where $\chi_b=\chi_b^0+\tilde\chi_b(\epsilon t)$ is the actual back position,
and write the level of $h$
at the back as $h_b=h(\chi_b)=h_b^0+\epsilon h_b^1(\epsilon t)$. Using these
forms in (2), assuming $h$ is constant across the back, we find
$$\partial_\chi^2\phi_b^1+c\partial_\chi\phi_b^1+(1-3{\phi_b^0}^2)\phi_b^1=
h_b^1-\epsilon^{-1}\dot\chi_b{d\phi_b^0\over d\chi},\eqno(7)$$
where the dot over $\chi_b$ denotes differentiation with respect to $t$.
Solvability of (7) requires the right hand side of that equation
to be orthogonal to $(d\phi_b^0/d\chi)\exp(c\chi)$. This leads to
$$\dot\chi_b=\beta\epsilon h_b^1,\eqno(8)$$
where $\beta$ is a positive constant. Using (5) to evaluate $h_b=h(\chi_b)$
and consequently $\epsilon h_b^1$ we find from (8)
$$\dot{\tilde\chi_b}=\beta(h_+-h_b^0)\bigl(1-e^{\epsilon\kappa\tilde\chi_b}
\bigr).\eqno(9)$$
The linearization of (9) about $\tilde\chi_b=0$ gives the equation
$\dot{\tilde\chi_b}=-\epsilon\kappa\beta(h_+-h_b^0)\tilde\chi_b$. Since
$\kappa>0$, we conclude that for $\epsilon\ll 1$, a stable localized structure
is formed whenever $h_+>h_b^0$.

We consider now the other extreme, $\epsilon\gg 1$.
Adiabatic elimination of $h$ reduces (2+4) to the variational
equation
$$\phi_t=(1-a_1^{-1})\phi-\phi^3+a_0/a_1+\phi_{xx}\eqno(10)$$
with $h=\phi/a_1-a_0/a_1$, where we  assumed that $\delta\ll\epsilon$.
Equation (10) is equivalent to (2) and, consequently,  has only one type of
front solution connecting the two states $(\phi_\pm,h_\pm)$.
When $a_0=0$ the system (2+4) has an odd symmetry about $(\phi,h)=(0,0)$.
The two states $(\phi_\pm,~h_\pm)$ are equally stable and the front that
connects them is stationary. When $\delta=0$ it becomes an exact solution
of (2+4) that exists for all $\epsilon$ values. Fig. $1a$ shows a phase
portrait of this front solution
in the ($\phi,h)$ plane. It amounts to a straight diagonal line,
$h=a_1^{-1}\phi$, connecting
the two states. We recall that for $\epsilon\ll 1$ equations (2+4)
admit two types of propagating solutions. One may
therefore expect to find a bifurcation
from  a single to multiple front solutions  as
$\epsilon$ is decreased [12].

To study this bifurcation we consider the symmetric model ($a_0=0$) with a
non-diffusive $h$ field and write a propagating front
solution, $\phi=\phi_p(x-ct),~~h=h_p(x-ct)$, as power series in $c$
$$\eqalign{\phi_p=\phi_s+c\phi_1+c^2\phi_2+...\cr
h_p=h_s+ch_1+c^2h_2+...,\cr}\eqno(11)$$
where ($\phi_s$,~ $h_s=a_1^{-1}\phi_s$) is the stationary front solution.
Expanding $\epsilon$ as well, $\epsilon=\epsilon_0+c\epsilon_1+
c^2\epsilon_2+...$, and using these expansions in (2+4) (with $a_0=\delta=0$)
we find solvability conditions, one at each order, that determine the
coefficients $\epsilon_0,~\epsilon_1,...$. Carrying out this perturbation
scheme to third order in $c$ we find $\epsilon_0=a_1^{-2}$, $\epsilon_1=0$
and $\epsilon_2< 0$. These results imply a supercritical pitchfork bifurcation
occurring at $\epsilon_c=\epsilon_0=a_1^{-2}$. Near the bifurcation the
front speed scales like $c\sim (\epsilon_c-\epsilon)^{1/2}$. Numerical
studies on (2+4) confirm these results. A numerically computed bifurcation
diagram for the symmetric case $a_0=0$ is shown in Fig. $2a$.

The leading order corrections in (11) take the form
$\phi_1=0,~~h_1=\phi_s^\prime$. Using these forms in (11) we see that the
difference between the stationary and the counter
propagating solutions close to the bifurcation point is that in the latter
the field $h$ is {\it translated} to the right or to the left by an amount
proportional to $c$. This translation breaks the odd symmetry of the
stationary solution and gives rise to phase portraits deviating from the
diagonal, $h=a_1^{-1}\phi$, as shown in Fig. $1b$. The speed $c$ can be
directly related to this deviation by using (11):
$$c=\alpha_s\int_{-\infty}^\infty\psi\phi_s^\prime d\chi~~~~~~~~~~
\psi=h_p-a_1^{-1}\phi_p,\eqno(12)$$
where $\alpha_s=1/\int_{-\infty}^\infty (\phi_s^\prime)^2 d\chi$ is positive.
Thus a front connecting $(\phi_+,h_+)$ at $\chi=-\infty$ to
$(\phi_-,h_-)$ at $\chi=\infty$ propagates to the right ($c>0$) when the
deviation from the diagonal is negative ($\psi<0$)
and to the left when the deviation is positive ($\psi>0$). As before,
we refer to the latter case as describing a back.

A significant distinction between the stationary front solution
$\phi=\phi_s,~~h= a_1^{-1}\phi_s$ (in the symmetric model) and the two
propagating solutions
$\phi=\phi_s,~~h= a_1^{-1}\phi_s\pm c\phi_s^\prime$ that bifurcate at
$\epsilon=\epsilon_c$ can be made by looking at the phase
$\theta=\arctan(h/\phi)$. Across the stationary front the phase
remains constant everywhere except for the core where
it suffers a jump of $\pi$ as the fields $\phi$ and $h$ vanish and change
sign. Across
a propagating front, on the other hand, the phase smoothly rotates by $\pi$
keeping the modulus $(\phi^2+h^2)^{1/2}$ nonzero. Similar types of walls
have been found in the context of anisotropic ferromagnets [18] and, recently,
in a nonvariational model describing a periodically forced oscillating
medium [19] (see also Refs. [20,21]). They are referred to as Ising walls when
the phase is singular and
as Bloch walls when the phase rotates smoothly.

The symmetric model is nongeneric unless the relevant physics dictates
an odd symmetry. Unfolding the symmetric case by allowing non zero values
of $a_0$ yields
a bifurcation diagram as shown in Fig. $2b$. In that case, front multiplicity
arises by a saddle node bifurcation occurring at
$\epsilon_{sn}(a_0)<\epsilon_c$.
For small $a_0$ we find the scaling
behavior, $\epsilon_c-\epsilon_{sn}\sim \vert a_0\vert^{1/2}$.

In the small $\epsilon$ regime we could show, using leading order singular
perturbation analysis, that front multiplicity gives
rise to persistent patterns. We
postpone the analogous analysis for higher $\epsilon$ values to a
subsequent study and present here, instead, results of direct numerical
integration of (2+4)
(with $\delta=0$). Our observations are summerized in a phase diagram
shown in Fig. 3 (see also Fig. 6 of Ref. [12]). To the right of the solid
curve $\epsilon=\epsilon_{sn}(a_0)$ only one type
of front solution exists and, as we expect, initial patterns decay toward
a uniform state. We note that along the line
$a_0=0$ the decay can be extremly slow and unnoticible in practical
situations [22]. As we cross the solid curve  by decreasing $\epsilon$
below $\epsilon_{sn}(a_0)$ front multiplicity arises. For some range of
$\epsilon$ initial patterns still decay toward a
uniform state. Persistent patterns, in the form of stable
solitary and periodic traveling waves, appear  below a second critical
value of $\epsilon$, that is, in the region to the left of the dashed curve.

We have presented here a mechanism of pattern formation in systems undergoing
a bifurcation from Ising to Bloch type fronts. A key ingredient in this
mechanism is the coexistence of stable, counter propagating front solutions
connecting the same asymptotic states as $\vert\chi\vert\to\infty$.
Coexistence of such fronts cannot occur in variational systems having free
energies. The mechanism can be tested
in bistable chemical reactions [2].


We would like to thank Rob Indik and Joceline Lega for interesting
discussions. We also
wish to thank the Arizona Center for Mathematical Sciences (ACMS)
for support. ACMS is sponsored by AFOSR contract FQ8671-9000589
(AFOSR-90-0021) with the University Research Initiative Program at the
University of Arizona. Part of this work has been done while A.H. was
visiting the Center for Nonlinear Studies at Los Alamos National Laboratory
and E.M. the Chemical Physics Department at the Weizmann Institute.
We thank these places for their warm hospitalities.

\vfil\eject
\noindent
{\bf References:}\hfil
\item{1.} E. Moses, J. Feinberg and V. Steinberg, Phys. Rev. A
{\bf 35}, 2757 (1987); R. Heinrichs, G. Ahlers and D. S. Cannell, Phys. rev.
A {\bf 35}, 2761 (1987); P. Kolodner, D. Bensimon and C. M. Surko, Phys.
Rev. Lett. {\bf 60}, 1723 (1988); K. E. Anderson and R. P. Behringer,
Phys. Lett. {\bf 145}, 323 (1990); H. Riecke, Phys. Rev. Lett. {\bf 68},
301 (1992).
\item{2.}  R. J. Field and M. Burger, {\it Oscillations and Traveling
Waves in Chemical Systems} (Wiley, New York, 1985). See also Refs. [8] below.
\item{3.} C. Elphick and E. Meron, Phys. Rev. A {\bf 40}, 3226 (1989);
B. Denardo, W. Wright and S. Putterman, Phys. Rev. Lett. {\bf 64},
1518 (1990).
\item{4.} O. Thual and S. Fauve, J. Phys. (Paris) {\bf 49}, 1829 (1988);
S. Fauve and O. Thual, Phys. Rev. Lett. {\bf 64}, 282 (1990).
\item{5.} V. Hakim, P. Jakobsen and Y. Pomeau, Europhys. Lett. {\bf 11},
19 (1990); Y. Pomeau, Physica D {\bf 51}, 546 (1991).
\item{6.} W. van Saarlos and P. C. Hohenberg, Phys. Rev. Lett. {\bf 64},
749 (1990); Physica D {\bf 56}, 303 (1992).
\item{7.} T. Ohta and M. Mimura, in {\it Formation, Dynamics and Statistics of
Patterns I}, eds. K. Kawasaki, M. Suzuki and A. Onuki (World Scientific,
New York, 1990).
\item{8.} J. J. Tyson and J. P. Keener, Physica D {\bf 32}, 327 (1988);
A. S. Mikhailov, {\it Foundation of Synergetics I. Distributed
Active Systems} (Springer, Berlin, 1990);
E. Meron, ``Pattern Formation in Excitable Media'', to appear in Physics
Reports.
\item{9.} P. C. Fife, CBMS-NSF Regional Conf. Series in Appl. Math. {\bf 53},
1 (1988).
\item{10.} P. Ortoleva and J. Ross, J. Chem. Phys. {\bf 63}, 3398 (1975).
\item{11.} P. C. Fife, J. Chem. Phys. {\bf 64}, 554 (1976).
\item{12.} J. Rinzel and D. Terman, SIAM J. Appl. Math. {\bf 42} 1111 (1982).
\item{13.} H. Ikeda, M. Mimura and Y. Nishiura, Nonl. Anal. TMA {\bf 13},
507 (1989).
\item{14.} D. G. Aronson and H. F. Weinberger in {\it Proceedings of the
Tulane Program in Partial Differential Equations and Related Topics}, ed.
J. Goldstein (Springer, Berlin, 1975).
\item{15.} J. L. Jewell, H. M. Gibbs, S. S. Tarng, A. C. Gossard, and W.
Weigmann, Appl. Phys. Lett. {\bf 40}, 291 (1982); E. Abraham, Optics
Communications {\bf 61}, 282 (1987).
\item{16.} J. B. Collins and H. Levine, Phys. Rev. {\bf B31}, 6119 (1985);
G. Caginalp, Ann. Phys. {\bf 172}, 136 (1986).
\item{17.} J. D. Murray, {\it Mathematical Biology} (Springer, New York,
1989).
\item{18.} J. Lajzerowicz and J. J. Niez, J. Phys (Paris) Lett. {\bf 40}
L165 (1979).
\item{19.} P. Coullet, J. Lega, B. Houchmanzadeh, and J. Lajzerowicz,
Phys. Rev. Lett. {\bf 65} 1352 (1990).
\item{20.} S. Sarker, S. E. Trullinger, and A. R. Bishop, Phys. Lett.
{\bf 59A}, 255 (1976).
\item{21.} P. Coullet,  J. Lega, and Y. Pomeau, Europhys. Lett. {\bf 15},
221 (1991).
\item{22.} K. Kawasaki and T. Ohta, Physica {\bf 116A}, 573 (1982);
J. Carr and R. L. Pego, Comm. Pure and Appl. Math. {\bf 42},
523 (1989).
\vfil\eject
\noindent
{\bf Figure Captions:}\hfil\break
{\sl Figure 1:}~ Phase portraits of front solutions connecting the
$(\phi_+,h_+)$ state at $\chi=-\infty$ to the  $(\phi_-,h_-)$ state at
$\chi=\infty$. The light colored curves are the nullclines $f=0$ and $g=0$
and the dark colored curves are the numerically computed trajectories.
The computational parameters are: a.~ $\epsilon=1.0,~ \delta=0,~ a_1=2.0,~
a_0=0$.~ b.~ $\epsilon=.2,~ \delta=0,~ a_1=2.0,~
a_0=0$. c. \hfil\break
{\sl Figure 2:}~ Bifurcation diagrams of front solutions. The dots are data
points representing  the speed of the different types of stable front solutions
that exist for each value of $\epsilon$. a.~ The symmetric case ($a_0=0$).
b.~ The nonsymmetric case ($a_0=.1$).
\hfil\break
{\sl Figure 3:}~ Phase diagram in the $\epsilon-a_0$ plane. For the region
to the right of
the solid curve only one type of front solution exists and initial patterns
do not persist. In the region between the solid and the dashed curves
multiple stable fronts
coexist but patterns still decay toward a uniform state. For the region to
the left of the dashed curve initial patterns evolve toward persistent
patterns in the form of stable traveling waves. Computational parameters
are :~ $\delta=0,~ a_1=2.0$.\hfil

\vfil
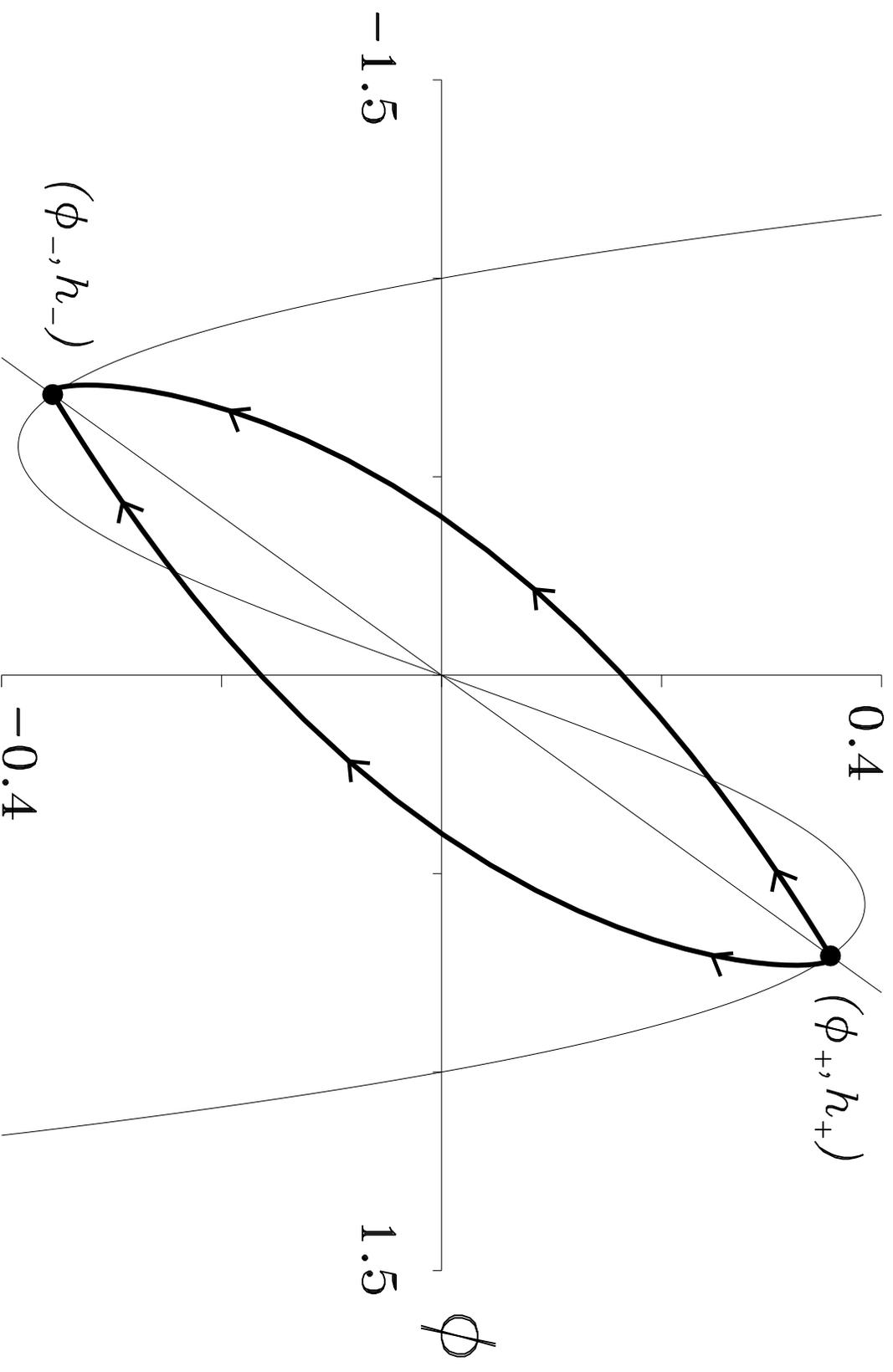
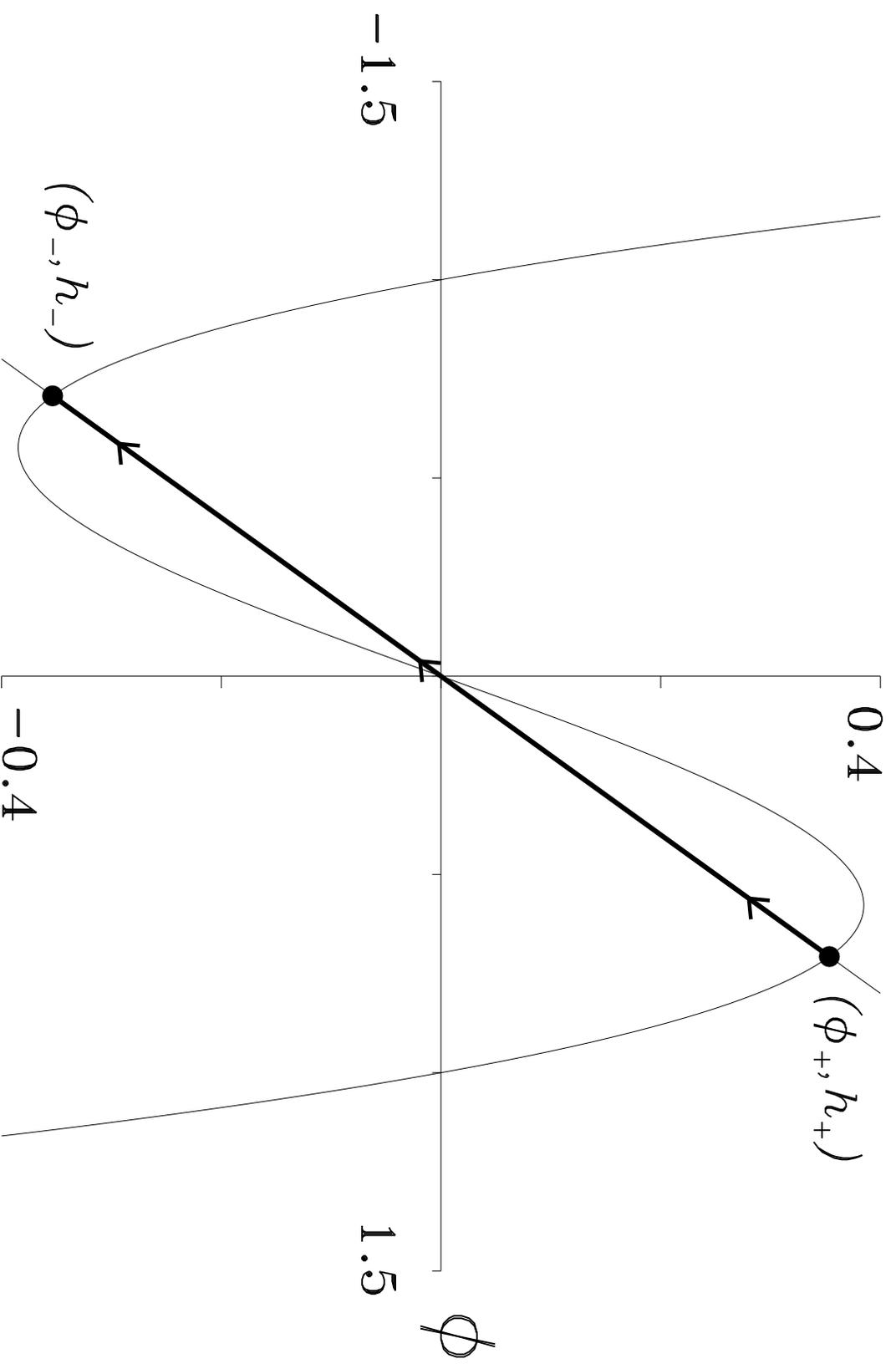
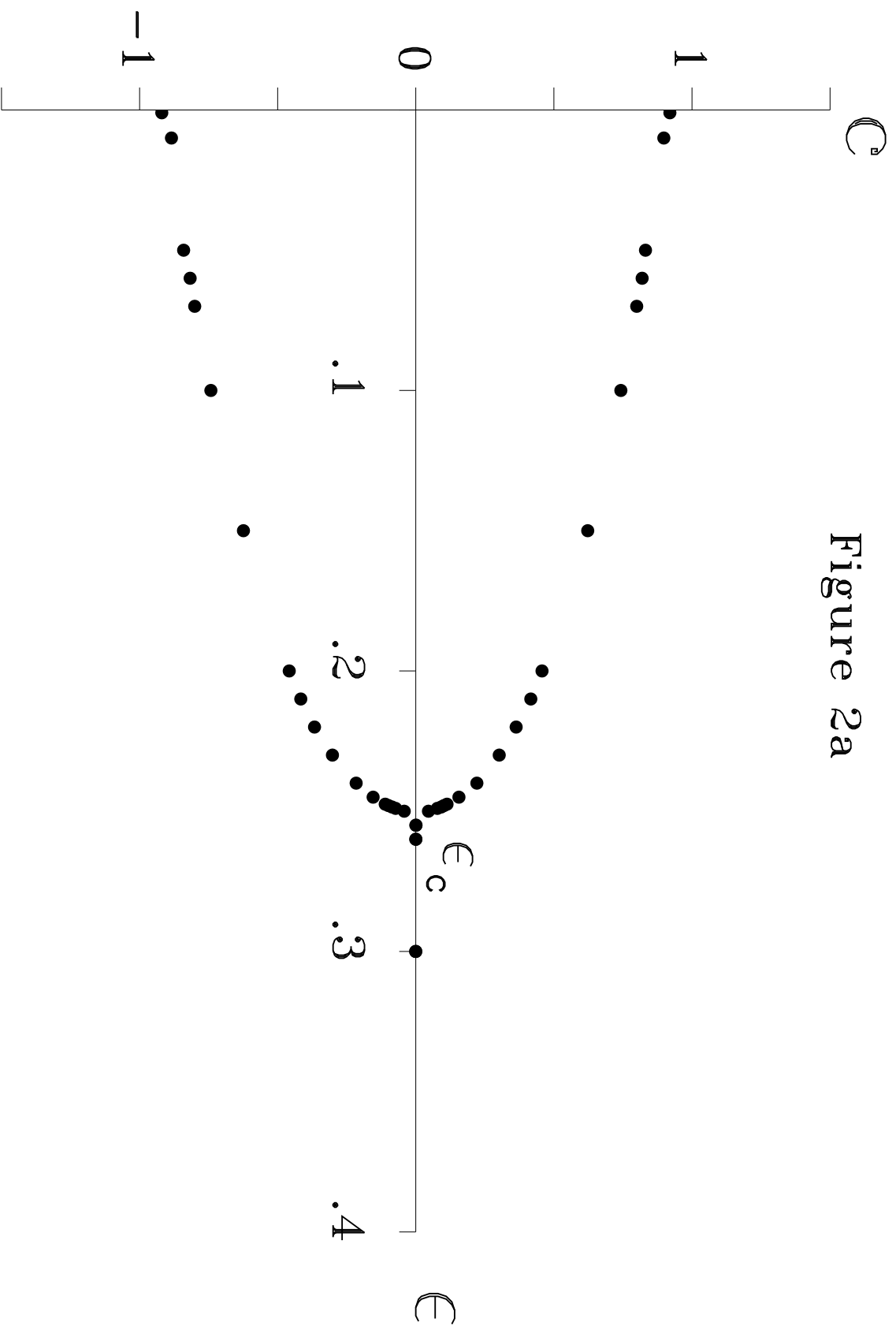
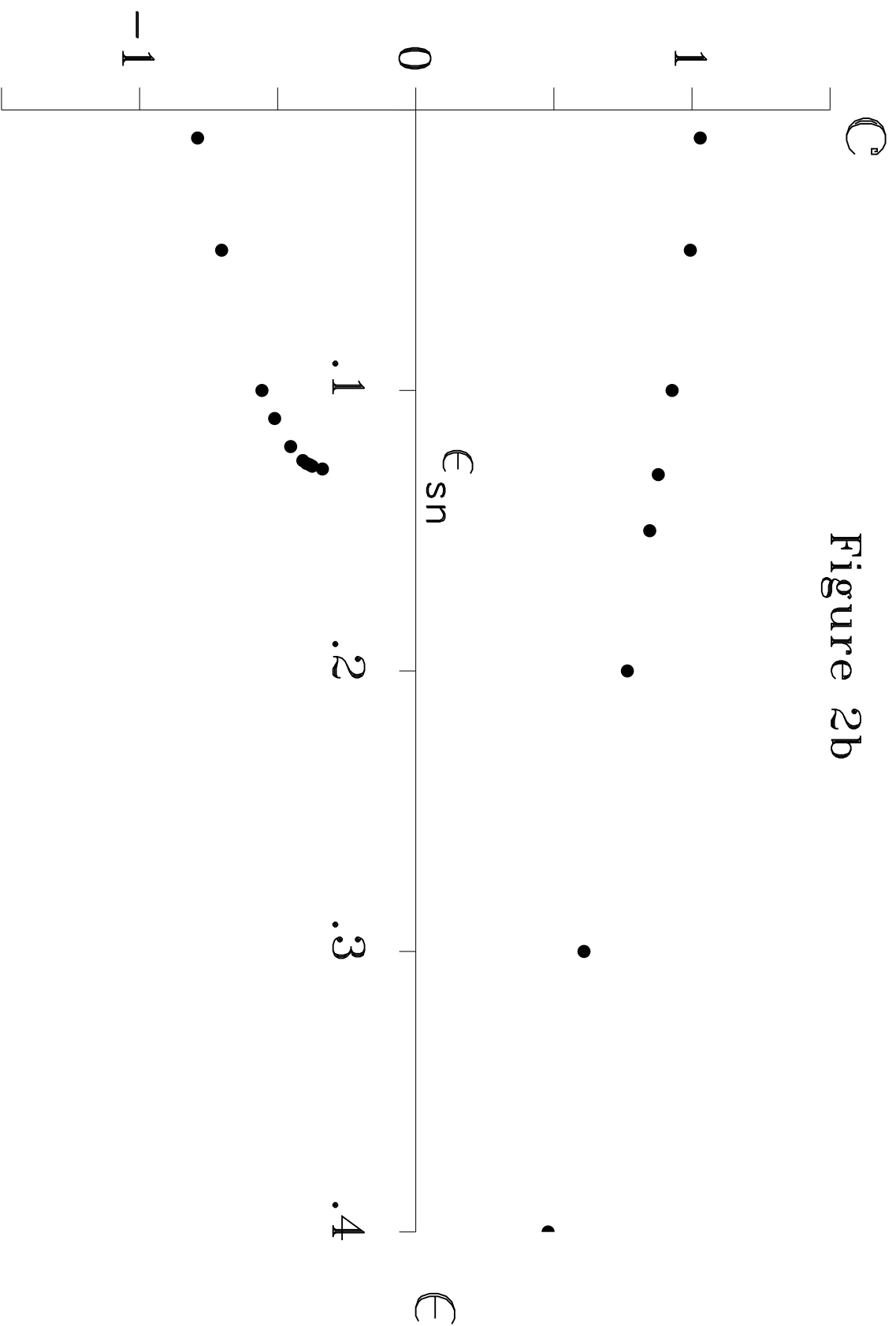
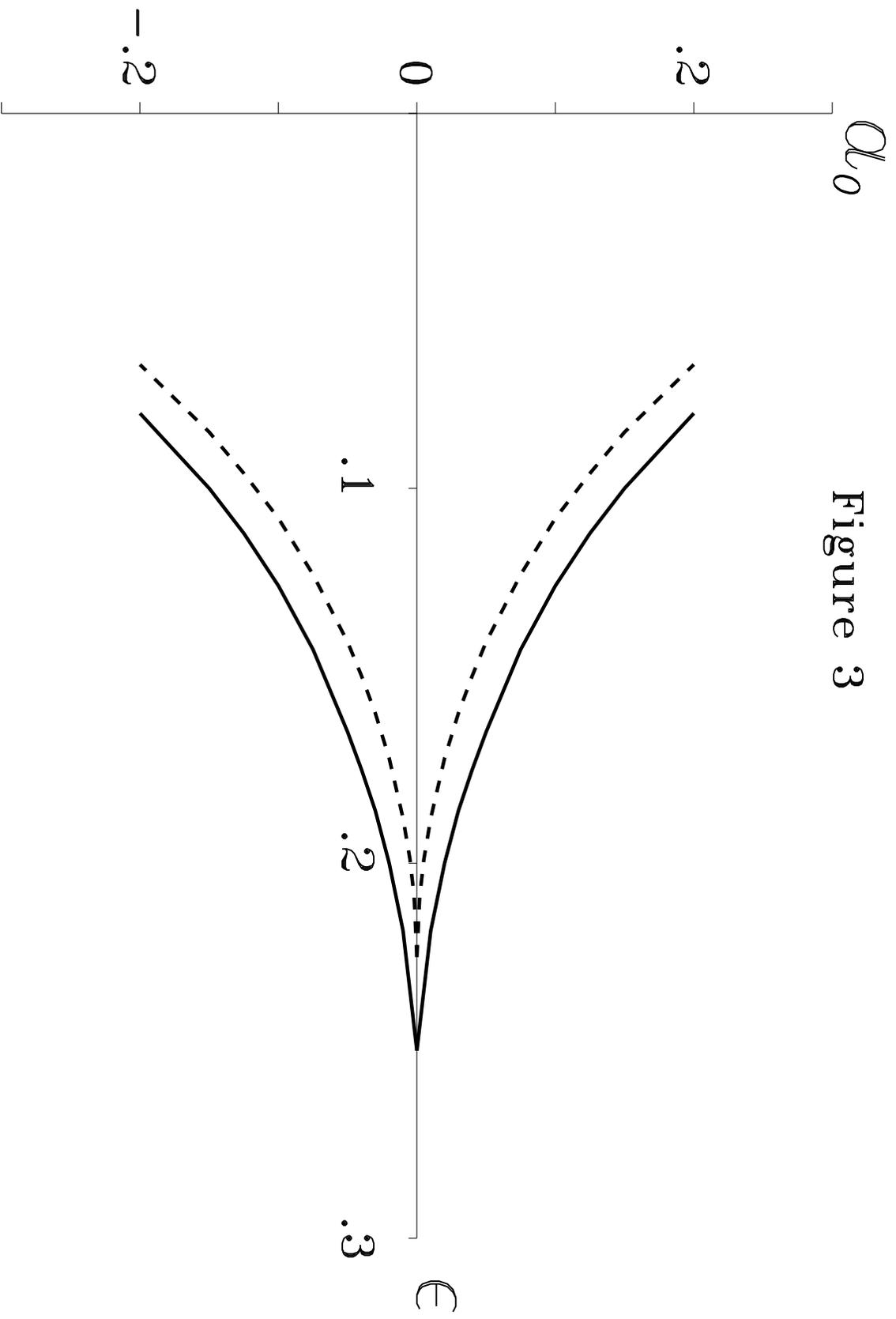
\bye